\documentclass[final,3p,times,twocolumn]{elsarticle}
\usepackage{lineno}
\usepackage{color}

\usepackage{amssymb}
\usepackage{subfigure}
\begin{document}
  \begin{frontmatter}
    \title{Optical and structural properties of CsI thin film photocathode}
    \author{Triloki, R. Rai}
    \author {B. K.~Singh\corref{cor}}
    \cortext[cor]{Corresponding author}
    \ead{bksingh@bhu.ac.in}
    \address{High Energy Physics laboratory, Department of Physics, Banaras Hindu University, Varanasi-221005, INDIA}

    \begin{abstract}
      In the present work, the performance of a cesium iodide thin film photocathode is studied in detail. The optical absorbance of cesium iodide films has been analyzed in the spectral range from 190 nm to 900 nm. The optical band gap energy of 500 nm thick cesium iodide film is calculated from the absorbance data using  Tauc plot. The refractive index is estimated from envelope plot of transmittance data using Swanepoel's method. The absolute quantum efficiency  measurement has been carried out in the wavelength range from 150 nm to 200 nm. The crystallographic nature and surface morphology are investigated by X-ray diffraction and transmission electron microscopy techniques. In addition, the elemental composition result obtained by energy dispersive X-ray analysis is also reported in the present work.
    \end{abstract}
    
  \end{frontmatter} 
  
  \section{Introduction}
  Photocathode devices in the soft X-ray and ultraviolet (UV) wavelength range are very important in the particle physics experiments for particle identification~\cite{NIMA_639_2011_117,NIMA_366_1995_345,NIMA_371_1996_155,NIMA_523_2004_345}. In soft X-ray and UV wavelength regions, alkali halide photocathodes are known to be very efficient photo converters. Cesium iodide (CsI) is one of the most efficient among them due to its highest quantum efficiency (QE) and good stability under short exposure to humid air~\cite{NIMA_504_2003_4}. Therefore it is widely used in many UV-detecting devices~\cite{IEEE_NS_48_3_2001,Proc_SPIE_59200I_1}. These devices consist films of thicknesses varying from few nanometer (nm) to micrometer ($\mu$m), depending upon the mode of operation and application of photocathode. It is very important to know the absorbance, transmittance and refractive index as a function of wavelength to predict the photoemissive behavior of a photocathode device. Knowledge of these optical constants is also necessary to determine the optical band gap energy.

  In the present work, the optical absorbance of CsI thin films is measured and analyzed. The optical transmittance and band gap energy are derived from the optical absorbance data of CsI thin films. The dispersive behavior of 500 nm thick CsI film is studied by calculating the refractive index. The value of refractive index has been determined by using Swanepoel's method. The photoemission properties of 500 nm thick CsI photocathode is studied in the spectral range from 150 nm to 200 nm. The structural, morphological and elemental composition analysis are also reported in the present work.
  
  \section{Experimental Details}
  
\begin{figure}[!ht]
    \begin{center}
      \includegraphics[scale=0.35]{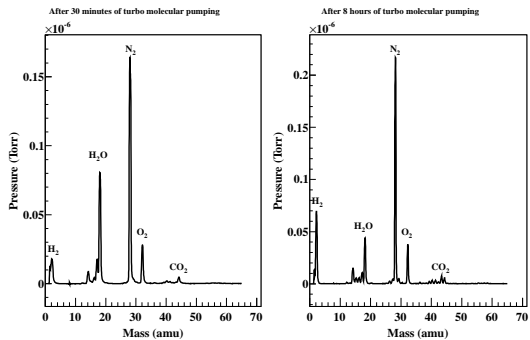}
      \caption{Residual gas composition inside the vacuum chamber: After 30 minutes of pumping (left panel) and after 8 hours of pumping (right panel).}
      \label{RGA_multigraph.eps}
    \end{center}
  \end{figure}

  The CsI thin films evaporation, used for present test, are carried out into a high vacuum stainless steel chamber. The evaporation chamber is pumped with a turbo-molecular pump (model: TMU 521 P, Pfeiffer) having a pumping speed of 510 L/s for $N_{2}$ gas. Prior to CsI evaporation, residual atmosphere of the chamber is monitored through a residual gas analyzer (model: SRS RGA 300), under a high vacuum environment ($3\times 10^{-7}$ ~Torr). It is observed that a large amount of water molecules has been evacuated from the evaporation chamber after 8 hours of pumping (see Figure \ref{RGA_multigraph.eps}). The main constituents of residual gases (at partial pressure $3\times10^{-7}$~Torr) are $N_{2}$ $(58.2\%),$ $H_{2}~(11.0\%),$ $H_{2}O~(16.2\%),$ $O_{2}~(9.9\%)$ and $CO_{2}~(1.4\%)$. After 8 hours of pumping, the majority of residual gases remaining inside the evaporation chamber is $N_{2}$. The $N_{2}$ gas does not affect the properties of CsI photocathode during film preparation. A small amount of CsI crystals (Alfa Aesar, 5N purity) are placed into a tantalum (Ta) boat inside the vacuum chamber and carefully heated to allow out gassing from its outer surface. After proper out gassing and melting, CsI thin films are deposited on quartz, aluminium (Al) and formvar coated copper (Cu) grid substrates. For uniform deposition, distance between Ta boat and substrate is kept at $\sim$20 cm. The CsI films are deposited at a typical rate of 1 nm to 2 nm per second.
 
  \begin{figure}[ht!]	
    \begin{center}
      \includegraphics[scale=0.25]{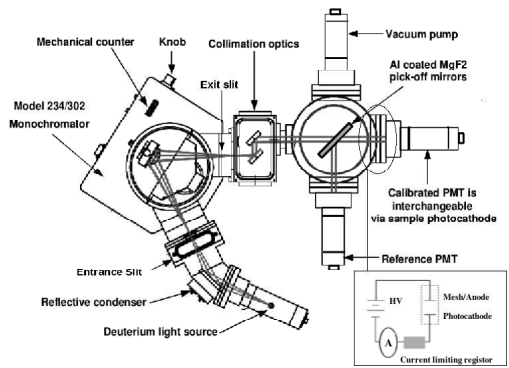}
      \caption{A schematic view of the absolute quantum effiency measurement setup, equipped with deuterium lamp, PMT and collimating optics.}
      \label{qen1.eps}
    \end{center}
  \end{figure}
  
 The thickness of the films are controlled by a quartz crystal thickness monitor (Sycon STM 100). After the sample preparation, vacuum chamber is purged with dry $N_{2}$ gas, in order to avoid the interaction of water vapor present in humid air with the prepared sample. Immediately after the chamber opening under constant flow of $N_{2}$ gas, CsI films are placed into a vacuum desiccator and further moved to characterization setup.

  The schematic diagram of experimental setup for the absolute quantum efficiency (QE) measurement is shown in Figure~\ref{qen1.eps}. This experimental setup includes a high vacuum stainless steel (S.S.) chamber (pumped with a TMP to $10^{-5}$~Torr), coupled to a vacuum ultra violet (VUV) monochromator (model: 234/302 VUV monochromator, McPherson). The VUV monochromator is equipped with a 30 watt magnesium fluoride ($MgF_{2}$) windowed deuterium $(D_{2})$ lamp (model: 632, McPherson) having spectral range from 113 nm to 380 nm. The QE of 500 nm thick CsI sample is measured in spectral range from 150 nm to 200 nm. A positive voltage is applied from high voltage power supply (CAEN - N471A) to a mesh electrode placed at a distance of $\sim$3 mm from the photocathode surface. The photocurrent, induced by monochromatic UV photons from photocathode surface, is recorded by a picoammeter (Keithley - 6485). The absolute QE, which is the ratio of number of emitted photoelectrons ($N_{e}$) to number of incident photons ($N_{p}$) i.e. $QE=N_{e}/N_{p}$, is derived from the ratio of the current measured from the photocathode to the current measured from a calibrated photomultiplier (Cal. PMT). The absolute QE measurement is done by alternatively directing the UV beam to both Cal. PMT or to the interchangeable photocathode. This PMT is calibrated against a National Institute of Standard and Technology (NIST) vacuum-photodiode. The stability of $D_{2}$ lamp is monitored throughout the measurements by a second reference photomultiplier (model: 658, Hamamatsu), of the same type and the measured photocurrent values are corrected correspondingly. 
  
  The optical properties measurement of CsI films are carried out using Perkin Elmer $\lambda 25$ UV/Vis spectrophotometer in the wavelength range from 190 nm to 900 nm. The structural properties of 500 nm thick CsI film is studied by using X'Pert PRO PANalytical X-Ray diffractometer. This X-Ray diffractometer is based on Bragg-Brentano para-focusing geometry. The diffractometer is operated at 30 kV and 40 mA with Cuk$\alpha$ ($\lambda$ = 1.54056 $\AA$) radiation. The morphological and structural features of a 500 nm thick CsI film deposited on formvar coated Cu grid are studied by transmission electron microscopy (TEM) of FEI Technai 20 $G^{2}$, operated at 200 kV accelerating voltage. The elemental composition of 500 nm thick CsI film is studied by the means of energy dispersive X-ray (EDAX) spectroscopy technique.

  \section{Optical properties of CsI thin films}
  
  \subsection{Optical absorbance}
  
\begin{figure}[ht!]
    \begin{center}
      \includegraphics[scale=0.35]{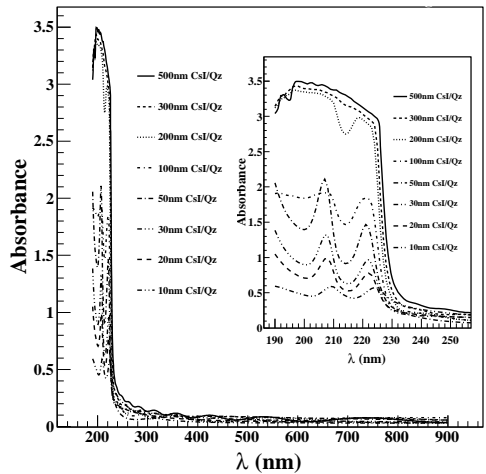}
      \caption{Optical absorption of CsI thin films of different thickness in the wavelength range from 190 nm to 900 nm and zoomed view of absorbance in UV spectral region (inset)}
      \label{Inset_abs.eps}
    \end{center}
  \end{figure}

  The UV/Vis absorption of CsI films deposited on quartz substrate are performed in the spectral range from 190 nm to 900 nm. For the optical absorbance measurement, the quartz substrate is chosen, because it is transparent in this spectral range. The absorption spectra of CsI films are shown in Figure \ref{Inset_abs.eps}. It is observed that the absorbance of CsI films varies between 0 to 2 for 10 nm to 100 nm thickness, while for thickness more than 100 nm, it lies between 0 to 3.5 (as shown in inset of Figure \ref{Inset_abs.eps}).

  Two strong absorption peaks are observed in the UV wavelength region at a wavelength smaller than 225 nm for thinner CsI films. A similar optical absorbance results for thinner and thicker CsI films are reported in previous literatures~\cite {NIMA_343_1994_135,NIMA_362_1995_183,JAP_83_1998_7896}.

  The value of absorption coefficient $(\alpha)$ is calculated from the absorption spectrum using the relation~\cite {NIMA_482_2002_238}:
  
  \begin{equation}
    \alpha = \frac{1}{t}ln\frac{1}{T},
  \end{equation}
  where t is the thickness of the film and T is the transmittance of the film. The absorption coefficient $(\alpha)$, estimated using equation (1), lies in between 0.02 to 0.04.
  
  \subsection{Optical transmittance}
  
\begin{figure}[ht!]
    \begin{center}
      \includegraphics[scale=0.35]{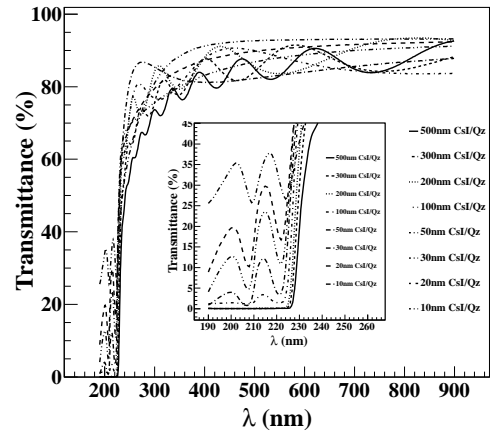}
      \caption{Optical transmittance of CsI thin films of different thickness in the wavelength range from 190 nm to 900 nm.}
      \label{inset_trans.eps}
    \end{center}
  \end{figure}

  The optical transmittance of CsI films are derived from the absorbance data in the wavelength range from 190 nm to 900 nm. The transmittance of CsI films is shown in Figure~\ref{inset_trans.eps}. Several transmittance peaks of CsI films are observed and are in good agreement with the previous reported work~\cite{NIMA_482_2002_238,NIMA_523_2004_323}. The optical transmittance  data are derived from the absorbance data using the relation~\cite {NIMA_343_1994_135}:

  \begin{equation}
    T = exp(-A),
  \end{equation}
  
  where A is the absorbance. 
  
  The transmittance result of CsI films for thicknesses more than 100 nm, depicts that they are opaque in the spectral region 190 nm to 225 nm and having a transmittance of about 2-3\% only. While, CsI films of thickness below 50 nm are found to be semitransparent in the spectral region 190 nm to 225 nm and varies from 20\% to 40\% (see inset of Figure ~\ref{inset_trans.eps}). A sharp increase in transmittance near a wavelength $\lambda\sim225$ nm is indicating its crystalline nature. Both of the thinner and thicker CsI films are found to be transparent in the spectral region of 225 nm to 900 nm and having more than 80\% transmittancy.
  
  The surface quality and homogeneity of CsI films are analyzed from the existence of interference fringes (oscillatory nature) in the transmittance spectra. In the transparent spectral region ($\lambda > 225$ nm), thinner and thicker CsI films show distinct characteristics related to inhomogeneities in the films. In this spectral region, CsI films of thickness smaller than 100 nm do not show any interference fringes pattern, which indicate that the CsI layers do not appear to be continuous and exhibiting small surface area coverage. While, for CsI films of thickness 100 nm and more, it show interference fringes pattern, which indicate the existence of continuous and homogeneous CsI layers, exhibiting large surface area coverage (see Figure~\ref{inset_trans.eps}). It is also observed from Figure~\ref{inset_trans.eps}, that oscillatory nature of CsI films increases with an increase in the thickness.

\begin{figure}[ht!]
    \centering
    \includegraphics*[width=70mm]{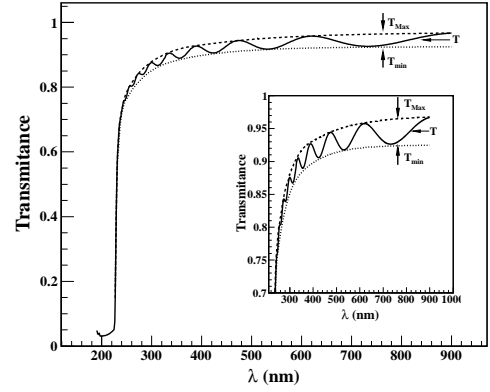}
    \caption{Transmission spectrum of 500 nm thick CsI thin film (solid line), including the maximum ($T_{Max}$) and minimum ($T_{min}$) transmittance envelope curves (dashed and doted lines).}
    \label{envelope.eps}
  \end{figure}

  The 500 nm thick CsI film is found to be more homogeneous and continuous than the thinner (semitransparent) CsI films. The transmission spectrum for 500 nm thick CsI film (shown in Figure~\ref{envelope.eps}) depicts a sharp fall in transmission near the fundamental absorption, which is an identification for the good crystallinity~\cite{MSP_25_2007_709,PSS_181_2000_427,PSS_106_1988_123,CRT_42_2007_275}. The oscillatory nature in the transmission spectrum observed for 500 nm thick CsI film is attributed to the interference of light transmitted through the thin film and the substrate.

  \subsection{Optical band gap energy of 500 nm thick CsI film}
  
  The optical band gap energy of a photocathode is one of the key parameter determining the range of its most efficient operation, in particular the sensitivity cutoff. In addition to proper band gap energy, a good photocathode material should allow an efficient electron transport to the emission surface and should have low or negative work function/electron affinity. 
  
\begin{figure}[ht!]
    \begin{center}
      \includegraphics[scale=0.35]{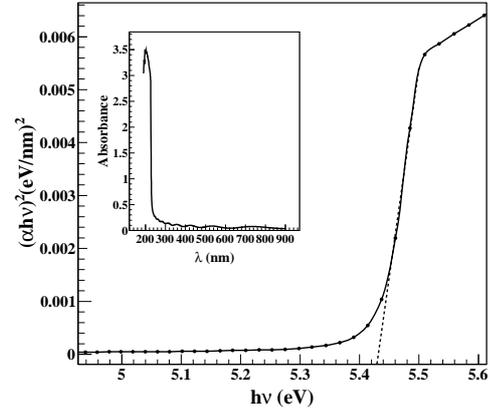}
      \caption{Variation of $(\alpha h \nu)^{2}$ vs. photon energy $h\nu$ and absorbance as a function of wavelength (inset) for 500 nm thick CsI film.}
      \label{eg3.eps}
    \end{center}
  \end{figure}

  The absorption in the UV wavelength region is attributed to band gap absorption of CsI thin film. An obvious increase in the absorption of wavelength less than 225 nm is observed (see inset of Figure ~\ref{eg3.eps}). It can be assigned to the intrinsic band gap absorption of CsI film due to the electron transmission from the valence band to conduction band. The absorption band gap energy ($E_{g}$) has been calculated by using the  Tauc relation \cite{PSS_15_1966_627,OPS_NH_1972,EPIC_OUPL_1940}:

  \begin{equation}
    (\alpha h\nu)^{n}=A(h\nu-E_{g}),
  \end{equation}

  where A is the edge width parameter, h is the Planck's constant, $\nu$ is the frequency of vibration, $h\nu$ is the photon energy, $\alpha$ is the absorption coefficient, $E_{g}$ is the band gap and n is either 2 for direct band transitions or 1/2 for indirect band transitions~\cite{JAP_87_2000_1318}. The direct optical band gap energy (shown in Figure ~\ref{eg3.eps}) is estimated from a Tauc plot of $(\alpha h \nu)^{2}$ versus photon energy $h\nu$ according to the K. M. model~\cite{JPC_97_1993_11802,TG_18_2011_117}. The value of photon energy $(h\nu)$ extrapolated to $\alpha$ = 0 gives an absorption edge which corresponds to a band gap energy $E_{g}$. The extrapolation gives a band gap energy of $E_{g}~\sim$5.4 eV which corresponds to absorption peak of 500 nm thick CsI film. The band gap energy determined from Tauc relation can be compared with band gap energy $E_{g}$ = 5.9 eV derived from experimental QE dependence on wavelength for heat-enhanced CsI thick film photocathode~\cite{JAP_77_1995_2138}.

  \subsection{Refractive index of 500 nm thick CsI film}
  
   The optical properties of 500 nm thick CsI film can be evaluated from transmittance data using Swanepoel method~\cite{JPSE_16_1980_1214,JPE_9_1976_1002}. The applicability of this method is limited to thin film deposited on transparent substrate. The application of this method entails, as a first step, the calculation of the maximum and minimum transmittance envelope curves by parabolic interpolation to the experimentally determined positions of peaks and valleys (shown in inset of Figure ~\ref{envelope.eps}). From maximum and minimum transmittance point, the values of refractive index ($n_{\lambda}$) are determined using the expression proposed by Swanepoel~\cite{JPSE_16_1980_1214} as given below:
  
  \begin{equation}
    n_{\lambda}=\sqrt{\left [N+\sqrt{N^{2}-n_{s}^{2}}\right]}.
\label{ref_ind_eqn}
  \end{equation}
  
  In the weak and medium absorption regions, the value of N is given by: 
  
  \begin{equation}
    N=2n_{s}\frac{T_{Max}-T_{min}}{T_{Max}T_{min}}+\frac{n_{s}^{2}+1}{2},
  \end{equation}
  
  with $n_{s}$ being the refractive index of the substrate. In general, $n_{s}$ is determined by the maximum of the transmission in the transparent region $T_{Max}$ ~\cite{AP_22_1989_199} using the relation:

  \begin{equation}
    n_{s}= \frac{1}{T_{Max}} + \sqrt{\left ( \frac{1}{T_{Max}^{2}}-1 \right )}.
  \end{equation}

 \begin{figure}[ht!]
    \begin{center}
      \includegraphics*[width=75mm]{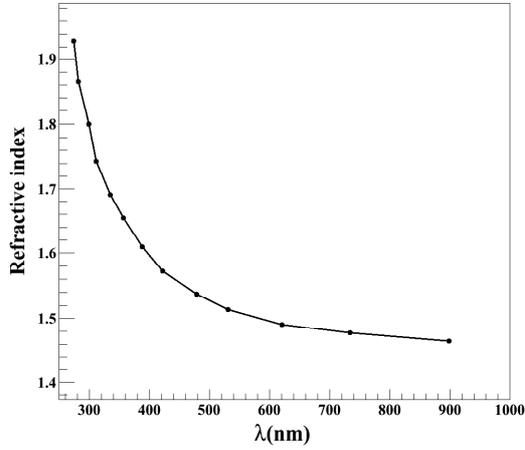}
      \caption{Refractive index as a function of wavelength for 500 nm thick CsI film deposited on quartz substrate.}
      \label{refrac.eps}
    \end{center}
  \end{figure}

  \begin{table}[ht]
    \begin{center}
      \caption[]{Values of $\lambda$, $T_{Max}$ and $T_{min}$ for the transmittance spectrum of CsI (Figure ~\ref{refrac.eps}) and the value of refractive index $n_{\lambda}$ to the corresponding wavelength $\lambda$:}
      \begin{tabular}{|l|l|l|l|}
	\hline
	$\lambda$ (nm) & $T_{min}$ & $T_{Max}$ & $n_{\lambda}$ \\ \hline
	275            & 0.81792 & 0.8455  & 1.930   \\
	283            & 0.83039 & 0.85802 & 1.860   \\
	300            & 0.85110 & 0.87923 & 1.800   \\
	312            & 0.86194 & 0.89029 & 1.740   \\
	336            & 0.87685 & 0.90815 & 1.690   \\
	359            & 0.88617 & 0.91966 & 1.655   \\
	390            & 0.89547 & 0.92872 & 1.610   \\
	423            & 0.90294 & 0.93507 & 1.572   \\
	479            & 0.91155 & 0.94488 & 1.554   \\
	532            & 0.91658 & 0.95134 & 1.513   \\
	622            & 0.92120 & 0.95850 & 1.489   \\
	735            & 0.92360 & 0.96364 & 1.478   \\
	900            & 0.92467 & 0.96731 & 1.464   \\
	\hline
      \end{tabular}
      \label{table_refraction} 
    \end{center}
  \end{table}

  It is observed that the value of refractive index decreases with an increase in wavelength, as shown in Figure ~\ref{refrac.eps}. Table 1 shows the values at the extremes of the spectrum of $\lambda$, $T_{Max}$ and $T_{min}$ obtained from envelope plot of Figure ~\ref{envelope.eps} (see zoomed view, inset image). The values of refractive index $n_{\lambda}$ calculated from equation ~\ref{ref_ind_eqn} are shown in Table~\ref{table_refraction}. The variation of the refractive index $n_{\lambda}$ with the wavelengths is shown in Figure~\ref{refrac.eps}. A sharp fall in refractive index is observed at the lower wavelength region and a gradual destruction is observed for refractive index corresponding to the higher wavelength region. This variation in refractive index indicates a normal dispersive behavior of 500 nm thick CsI film.

  \subsection{Photoemission properties of 500 nm thick CsI film}
  
  The photoemission properties of 500 nm thick CsI film deposited on Al substrate is studied in wavelength range from 150 nm to 200 nm, with a scan step size of 2 nm. The absolute QE, which is the ratio of emitted photoelectrons to incident photons, is determined by illuminating the CsI surfaces with photon flux of a given frequency and the resulting photocurrent is measured by a picoammeter. The current, the observable quantity,  is related to the QE by the following relation:
  
  \begin{equation}
    QE(\%) = \frac{I_{pc}}{I_{pm}}\times QE_{pm}\times G_{pm},
  \end{equation}
  
  where, $I_{pm}$ is the PMT current, $I_{pc}$ is the photocathode current, $QE_{pm}$ is known QE of Cal. PMT and $G_{pm}$ is gain of PMT.

  \begin{figure}[ht!]
    \begin{center}
      \includegraphics[scale=0.42]{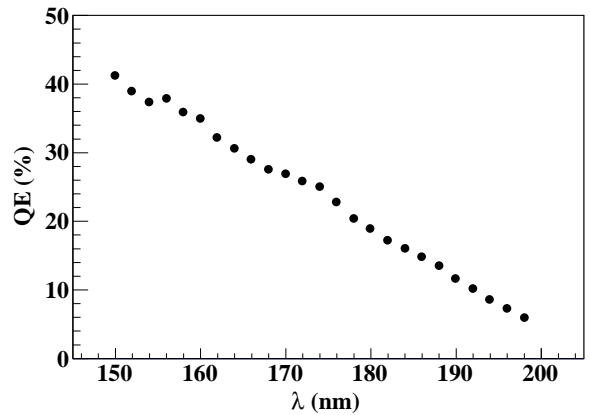}
      \caption{The absolute quantum efficiency (QE) as a function of wavelength for 500 nm thick thin CsI film deposited on Al substrate.}
      \label{QE1.eps}
    \end{center}
  \end{figure}

  It is clearly observed that from the plot of Figure~\ref{QE1.eps}, the maximum QE achieved is $\sim$40\% at a wavelength of 150 nm. The QE is found to decrease with an increase in wavelength of incident photon. The value of experimentally determined QE is in good agreement with the most of literature data for the CsI photocathode ~\cite{NIMA_581_2007_651,NIMA_695_2012_279}.

  \section{Structural properties of 500 nm CsI film}
  
  \subsection{Crystallographic analysis}
  
\begin{figure}[ht!]
    \begin{center}
      \includegraphics*[width=80mm]{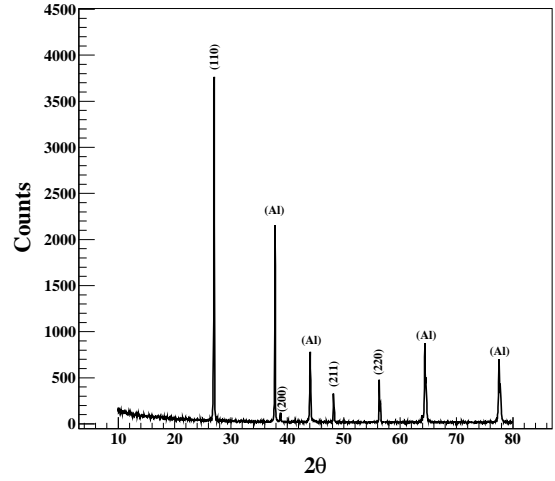}
      \caption{XRD pattern of 500 nm thick CsI film deposited on Al substrate. The diffraction pattern is taken in a Bragg Brentano parafocusing geometry.}
    \label{xrd.eps}
    \end{center} 
  \end{figure}

  The crystal structure and orientation of the 500 nm thick CsI film is investigated by X-ray diffraction (XRD). Figure~\ref{xrd.eps} shows the typical XRD pattern of 500 nm thick CsI film deposited on Al substrate. The XRD pattern indicates that, the CsI film is purely crystalline in nature. The XRD pattern contains an intense peak at Bragg's angle $2\theta = 27.06$, assigned to (110) crystallographic plane and three other XRD peaks are also found at Bragg's angles $2\theta = 38.87, ~48.30~ $and $~56.49$ corresponds to (200), (211) and (220) crystallographic planes, respectively. As CsI is deposited on Al substrate, so XRD pattern also contains four Al peaks (see Figure~\ref{xrd.eps}). These crystallographic planes attributed a body centered cubic (bcc) structure.
  
  The lattice constant (a) of crystalline CsI film is calculated using the analytical relation (for cubic crystal system)~\cite{XRD_PH_2001_3rd}:
  \begin{equation}
    a=d\times\sqrt{(h^{2}+k^{2}+l^{2})},
  \end{equation}
  where d is the interplanar spacing and (hkl) is the Miller indices of a plane. The value of lattice constant $(a)$ for 500 nm thick CsI thin film is found to be $\sim$4.66~\AA, which is in good agreement with lattice constant reported in International Center for Diffraction Data (ICDD, File number - 060311).
  
  The crystallite size of 500 nm thick CsI film is calculated using a well known Scherrer's equation~\cite{XRD_PH_2001_3rd,Gottinger_Nachrichten_2_1918_98}:

  \begin{equation}
    D=\frac{k\lambda}{\beta cos\theta},
  \end{equation}
  
  where D is the crystallite size, k (=0.9) is the crystal constant, $\lambda~(=1.5406\AA)$ is the wavelength of X-ray used, $\beta$ is the broadening of diffraction line measured at half of its maximum intensity in radians and $\theta$ is the angle of diffraction. The crystallite size obtained for most intense (110) crystallographic plane of CsI thin film is $\sim$55 nm, which matches very well with previous reported articles~\cite{NIMA_493_2002_16,NIMA_610_2009_350,NIMA_736_2014_128}
  
  Table 2 shows the values of Bragg's angle (2$\theta$), interplanar spacing (d), full width at half maximum (FWHM), texture coefficient $(TC_{(hkl)})$ and crystallite size (D), corresponding to their lattice planes (hkl) for 500 nm thick CsI film.

  \begin{table*}[ht]
    \begin{center}
      \caption[]{Structural parameters of a 500 nm thick CsI film analysed from XRD:}
    \vskip 0.1cm
      \begin{tabular}{|p{1.0 cm}|p{1.0cm}|p{1.0cm}|p{1.0cm}|p{1.0cm}|p{1.0cm}|p{1.0cm}|} \hline
 	(hkl)& $2\theta$[deg]& d[$\AA$]& FWHM& $\frac{I_{(hkl)}}{I_{0(hkl)}}$& $TC_{(hkl)}$& D(nm) \\ \hline
	
	(110)& 27.06& 3.295 & 0.1476& 1.00& 1.59& 55 \\ 
	
	(200)& 38.87& 2.315 & 0.1968& 0.37& 0.59& 43 \\ 
	
	(211)& 48.30& 1.884 & 0.1968& 0.45& 0.72& 44 \\
	
	(220)& 56.49& 1.628 & 0.2460& 0.69& 1.10& 46 \\
	\hline
      \end{tabular}
    \end{center} 
  \end{table*}

  \begin{figure}[ht!]
    \begin{center}
      \includegraphics[scale=0.32]{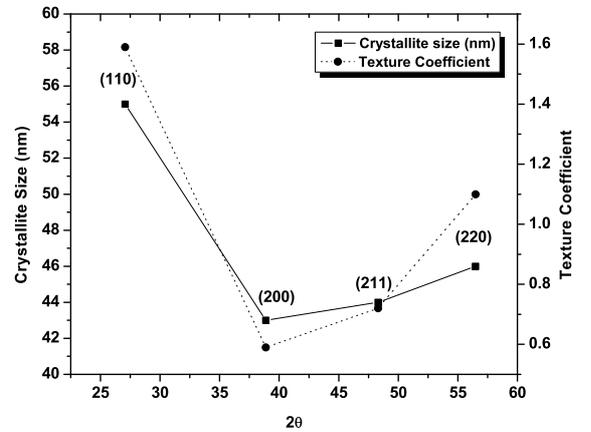}
      \caption{Average crystallite size and texture coefficient of 500 nm thick CsI thin film calculated from XRD pattern.}
      \label{tc.eps}
    \end{center}
  \end{figure}

  The XRD pattern shows a highly intense peak at $2\theta = 27.06$ indicating a strong preferred orientation along the (110) plane. The texture coefficient $TC_{(hkl)}$ of CsI determined from XRD data represents the texture of a particular plane. The deviation in $TC_{(hkl)}$ from unity implies the preferred growth. The $TC_{(hkl)}$ factor can be calculated for each crystallite orientation using the following equation~\cite{Structure_of_metal_PP_Oxford_1980}:
  
  \begin{equation}
    TC_{(hkl)}=\frac{I_{(hkl)}/I_{0(hkl)}}{N^{-1}\sum_{N} I_{(hkl)}/I_{0(hkl)}}\times100\%,
  \end{equation}
  
  where $I_{(hkl)}$ is the measured relative intensity of a plane (hkl) and $I_{0}(hkl)$ is the standard intensity of the plane (hkl) taken from the ICDD (card no: 060311) data and `N' is the number of reflections taken into account.
  
  By using the above equation, the preferred orientation of the lattice plane can be understood. The $TC_{(hkl)}$ is expected to be unity for the films with randomly oriented crystallites, while higher values indicate the abundance of grains oriented in a given crystallographic direction [hkl]. The variation of $TC_{(hkl)}$ and crystallite size for the peaks of the CsI film is presented in Table 2. $TC_{(hkl)}$ and average crystallite size of different planes of CsI film are shown in Figure~\ref{tc.eps}. The crystallite size varying from 43 nm to 55 nm and its average value is found to be $\sim$47 nm. 
 It is clear from the plot that the average pertaining to (110) reflection is higher than the other planes. This indicates that the preferential orientation of 500 nm thick CsI film corresponds to (110) crystallographic plane.

  \subsection{Morphological analysis}
  
  A detailed study on morphology of the reflective and semitransparent CsI thin film is reported in articles~\cite{NIMA_695_2012_279,NIMA_736_2014_128,NIMA_438_1999_409}. Previous results show that the reflective films having homogeneous and continuous grain like morphology along with larger value of grain size compared to semitransparent films. The impact of VUV irradiation and exposure to humid air on CsI surfaces are reported in references~\cite{NIMA_695_2012_279,NIMA_610_2009_350}. With the VUV irradiation and exposure to humid air, semitransparent CsI surfaces degrades quickly than the reflective film and its morphology becomes inhomogeneous and discontinuous.
   
\begin{figure}[ht!]
    \begin{center}
      \includegraphics[scale=0.25]{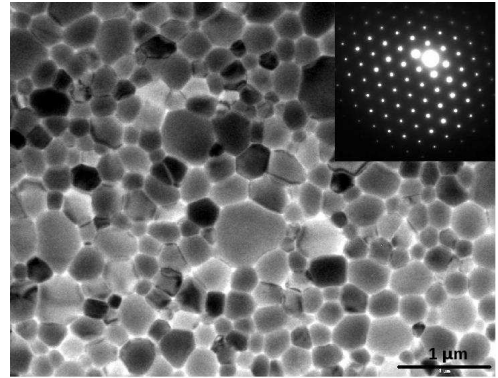}
      \includegraphics[scale=0.40]{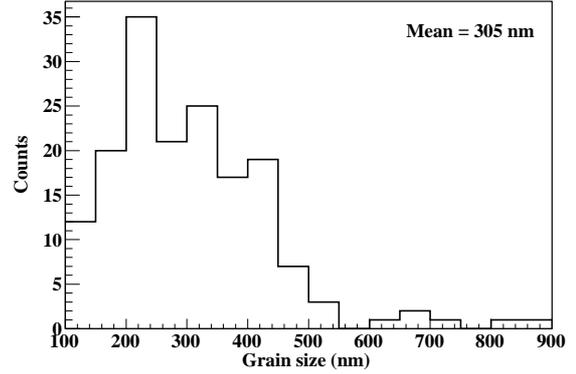}
      \caption{Transmission electron microscopy image and diffraction pattern (inset) of 500 nm CsI thin film (Top) and Grain size distribution of 500 nm CsI thin film (Bottom).} 
      \label{tem2_hist}
    \end{center}
  \end{figure}

  In this work, we have shown surface morphology of 500 nm thick reflective CsI film. The morphology of CsI surface is studied by TEM technique. A TEM is a powerful microscope that produces a high-resolution, black and white image from the interaction that takes place between samples and energetic electrons in the vacuum chamber. In order to observe the surface morphology of CsI film, few regions are scanned, one of them is shown in Figure~\ref{tem2_hist} (top panel). It is observed that the CsI film has homogeneous and continuous grain like morphology, with more than 95\% surface area coverage. The CsI film has grains of various sizes, ranging from 110 nm to 860 nm and average grain size is found to be $\sim$300 nm, as shown in histogram of Figure~\ref{tem2_hist} (bottom panel). The electron diffraction pattern shown in inset of Figure ~\ref{tem2_hist} (top panel) indicate that it is crystalline in nature and having single crystal like domains. A body centered cubic (bcc) structure with lattice constant $(a)$ $\sim$4.66~\AA ~is found to have a good match between experimental and calculated values of interplanar distances.

  \subsection{Elemental composition analysis of 500 nm thick CsI film}
  
  \begin{figure}[ht!]
    \begin{center}
      \includegraphics[scale=0.35]{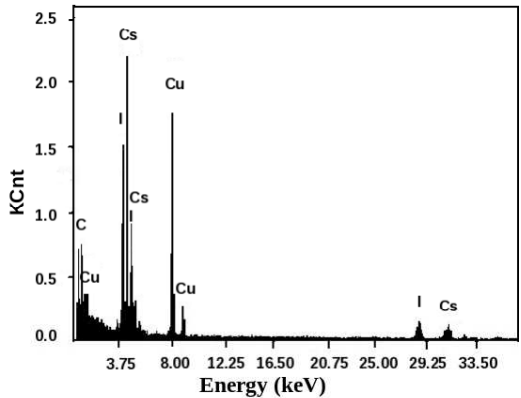}
      \caption{Chemical composition of a 500 nm thick CsI film is determined by EDAX. The spectra suggests that grains have chemical composition of Cs and I having Cs:I ratio is 1:1.}
      \label{edax1.eps}
    \end{center}
  \end{figure}
  
  The EDAX is used to detect elements present in considerable amount (quantitative determination of bulk element composition) of thermally evaporated CsI film. 
The elemental composition measurement of 500 nm thick CsI film, deposited on formvar coated Cu grid is performed by TEM system (shown in Figure~\ref{edax1.eps}). 
The data are shown with no smoothing, filtering or processing of any kind. The EDAX spectrum shows clear peaks corresponding to the  55 Cs K (30.97 keV) line, 55 Cs L line (4.23 keV), 53 I K line (28.51 keV) and 53 I L line (3.94 keV). The CsI film exhibits the cesium (Cs) and iodine (I) elemental peaks with an atomic ratio of Cs and I is 51:49 (thus the Cs:I ratio is found to be $\sim$1:1), which is consistent with the stoichiometry of CsI. The observed 29 Cu K line (8 keV), 29 Cu L line (1.5 keV) and 6 C K line (1 keV) peaks in the EDAX spectrum seen to be originated from the copper grid that is used for mounting the sample in the TEM machine. No other peaks are observed over the entire 0 keV to 35 keV detection windows.

  \section{Conclusions}
    The optical properties measurements of CsI thin films deposited on quartz substrate are performed in the spectral range from 190 nm to 900 nm. Two strong absorption peaks is observed in the UV wavelength region, one is at $\sim$207 nm and another is at $\sim$222 nm. The transmittance result of CsI films for thicknesses more than 100 nm, depicts that they are opaque in the spectral region 190 nm to 225 nm and having transmittance is about 2-3\%. The CsI films of thickness below 50 nm, found to be semitransparent in the spectral region 190 nm to 225 nm, where transmittance varies from 20\% to 40\%. A sharp increase in transmittance is observed near a wavelength $\lambda\sim$225 nm, indicating its crystalline nature. The thinner and thicker films are found to be transparent in the spectral region 225 nm to 900 nm and having more than 80\% transmittancy.

  Appearance of interference fringes pattern for 500 nm thick CsI film in transparent spectral region indicates, existence of continuous and homogeneous grain like morphology with maximum surface area coverage. The optical band gap energy has been calculated from the absorbance data using the Tauc plot of K. M. model. The optical band gap energy is found to be $\sim$5.4 eV. The values of refractive index calculated from envelop plot of transmittance data and it varies from 1.93 to 1.46 in the spectral range from 275 nm to 900 nm. This variation of refractive index indicating the dispersive behavior of CsI film.

  The photoemissive properties of thermally evaporated 500 nm thick CsI film has been investigated in the wavelength range from 150 nm to 200 nm. The maximum quantum efficiency achieved is $\sim$40\% at the wavelength $\lambda$ = 150 nm.
  
  Surface morphology, bulk structure as well as crystallographic nature of 500 nm thick CsI are studied by TEM and XRD techniques. TEM results reveal that it has homogeneous and continuous grain like morphology, with more than 95\% surface area coverage. The average grain size (composed of many coherent domains) of CsI film, obtained from TEM micro structure is $\sim$300 nm. The diffraction pattern obtained from XRD and TEM measurements reveals that CsI film is purely crystalline in nature and is having body centered cubic (bcc) structure. The observed value of lattice constant is obtained  $\sim$4.66~\AA. The average value of coherently scattering domain size (crystallite size) calculated using Scherrer's method is found to be $\sim$47 nm. The EDAX result indicates that, CsI film is having mainly Cs and I elements. The atomic ratio of Cs and I is found to be $\sim$1:1, which is consistent with the stoichiometry of CsI.

  \section*{Acknowledgments}
  This work is partly supported by the Department of Science and Technology (DST), the Council of Scientific and Industrial Research (CSIR) and by the Indian Space Research Organization (ISRO), Govt. of India. Triloki acknowledges the financial support obtained from UGC under Research Fellowship Scheme for Meritorious Students (RFSMS) program. R. Rai acknowledges the financial support obtained from UGC under research fellowship scheme in central universities.

  
  
  

\end{document}